\documentclass[preprint,showkeys,showpacs]{revtex4-1}
\usepackage{graphicx}
\begin{document}

\newcommand{\baru}{Ba(Fe$_{1-x}$Ru$_x$)$_2$As$_2$ }

\title{Physical and magnetic properties of \baru  single crystals}

\author{A.Thaler}
\author{N. Ni}
\author{A. Kracher}
\author{J. Q. Yan}
\author{S.L. Bud'ko}
\author{P.C. Canfield}
\affiliation{Ames Laboratory and Department of Physics and Astronomy, Iowa State University, Ames, IA 50011, USA}

\date{\today}

\begin{abstract}
Single crystals of \baru, $x<0.37$, have been grown and characterized by structural, magnetic and transport measurements.  These measurements show that the structural/magnetic phase transition found in pure BaFe$_2$As$_2$ at 134~K is suppressed monotonically by Ru doping, but, unlike doping with TM=Co, Ni, Cu, Rh or Pd, the coupled transition seen in the parent compound does not detectably split into two separate ones.  Superconductivity is stabilized at low temperatures for $x>0.2$ and continues through the highest doping levels we report.  The superconducting region is dome like, with maximum T$_c$ ($\sim16.5$~K) found around $x\sim 0.29$. A phase diagram of temperature versus doping, based on electrical transport and magnetization measurements, has been constructed and compared to those of the Ba(Fe$_{1-x}$TM$_x$)$_2$As$_2$ (TM=Co, Ni, Rh, Pd) series as well as to the temperature-pressure phase diagram for pure BaFe$_2$As$_2$.
%Because Ru doping does not provide extra electrons,
Suppression of the structural/magnetic phase transition as well as the appearance of superconductivity 
%must be a result of a combination of impurity effects directly from the doping and changes in unit cell dimensions.  This is in contrast to the Rh/Co and Pd/Ni phase diagrams, which by their similarity suggest extra electrons are much more important than unit cell dimensions or ratios.
is much more gradual in Ru doping, as compared to Co, Ni, Rh and Pd doping, and appears to have more in common with BaFe$_2$As$_2$ tuned with pressure; by plotting $T_S/T_m$ and $T_c$ as a function of changes in unit cell dimensions, we find that changes in the $c/a$ ratio, rather than changes in $c$, $a$ or V, unify the $T(p)$ and $T(x)$ phase diagrams for BaFe$_2$As$_2$ and \baru respectively.
\end{abstract}



\maketitle

\section{introduction}
The discovery of superconductivity in F-doped LaFeAsO\cite{kamihara} and~K-doped BaFe$_2$As$_2$\cite{rotter:107006} in 2008 led to extensive interest in these families of FeAs-based compounds.  The superconducting critical temperature, $T_c$, has risen as high as 56~K for F doped RFeAsO\cite{zhi-an} and as high as 38~K in K and Na doped AEFe$_2$As$_2$ systems (AE=Ba, Sr, Ca)\cite{rotter:107006}.  Superconductivity was also found in Co doped AEFe$_2$As$_2$\cite{sefat:117004} and RFeAsO\cite{sefat:104505}.  More recently, superconductivity has been found in other 3d, $4d$ and $5d$ transition metal, electron doped BaFe$_2$As$_2$ systems\cite{thaler,sharma,wang,saha,tillman,chu,ning,fang,canfield3d}, as well as SrFe$_2$As$_2$ and CaFe$_2$As$_2$.  Although the electron doped AEFe$_2$As$_2$ systems have lower $T_c$ values than the hole doped ones\cite{tillman,chu,ning,fang,canfield3d}, they have been studied extensively because doping is more homogeneous in these systems and single crystals can be more easily and reproduceably grown.  In order to understand the conditions for superconductivity in these systems, temperature versus doping phase diagrams must first be constructed.  Detailed studies have been made for TM doped BaFe$_2$As$_2$ (TM=Co, Ni, Cu, Rh, Pd, Pt, Ir)\cite{tillman,chu,ning,fang,lester,pratt,canfield3d,thaler}.  For Co, Ni, Cu, Rh and Pd, temperature vs doping concentration, $x$, and temperature vs electron count, $e$, phase diagrams show similar properties, with the temperature of the structural/magnetic transition, $T_S/T_m$, seen in the parent compound being suppressed and separated in a similar manner with $x$, and $T_c$ evolving in a similar manner with $e$, especially on the overdoped side of the superconducting dome\cite{thaler,tillman,canfield3d,canfieldoverview}.  Although TM doping of the BaFe$_2$As$_2$ system is convenient -- providing large homogeneous crystals -- it is not unique in tuning $T_S/T_m$ and $T_c$.  Pressure can also be used to suppress $T_S/T_m$ and stabilize a low temperature superconducting state\cite{torikachvili,alireza,colombier}.

In contrast with its 4d neighbors Rh and Pd, Ru doping provides no extra electrons to the bands.  However, recent polycrystalline studies in both the SrFe$_2$As$_2$\cite{schnelle,qi} and BaFe$_2$As$_2$\cite{sharma} systems show that Ru substitition on the Fe site suppresses the structural/magnetic phase transition and leads to superconductivity, indicating that this system may allow a direct comparison of nominally isovalent doping and electron doping TM substitution as well as pressure studies.  Isovalent doping induced superconductivity, as pressure before it, indicates that whereas $x$ and $e$ are important parameters in parameterizing the phase transitions in these systems, changes in the unit cell parameter may be important as well.

Based on this, we have studied Ru doped BaFe$_2$As$_2$ single crystals in order to compare the effects of isoelectronic doping to 3d and 4d transition metal, electron doped compounds.
%Single crystals are superior to polycrystalline samples for the determination of in plane versus out of plane inhomogeneity, as well as for electronic structure and inelastic scattering measurements.
As we wrote this work up, a similar, complimentary, study was posted;\cite{rullier,brouet} comparison to these data will be made as well.

\section{experimental methods}
Single crystals of \baru were grown out of self flux using conventional high-temperature solution growth techniques\cite{tillman,fisk}.  FeAs and RuAs were synthesized in the same manner as in \cite{tillman}.  Small Ba chunks and FeAs/RuAs powder were mixed together in a ratio of Ba:TMAs=1:4.  The mixture was then placed in an alumina crucible with a "catch" crucible filled with quartz wool placed on top.  Both crucibles were sealed in a silica tube under 1/6 atmosphere of Ar gas.  The sealed tube was heated up to 1180$^{\circ}$C over 12 hours, held at 1180$^{\circ}$C for 8-12 hours, and then cooled over 45-65 hours.  The final temperature varied between 1050$^{\circ}$C and 1100$^{\circ}$C, increasing with the Ru doping level. Once the furnace reached the final temperature, the excess FeAs/RuAs liquid was decanted, leaving the single crystals behind.  Unfortunately, this increasing decanting temperature made doping levels above $x=0.37$ difficult to produce.

Powder x-ray diffraction measurements, with a Si standard, were performed using a Rigaku Miniflex diffractometer with Cu $K\alpha$ radiation at room temperature. Diffraction patterns were taken on ground single crystals from each batch. Only very small FeAs impurity peaks were found as a secondary phase. The unit cell parameters were refined by "Rietica" software.  Elemental analysis of single crystal samples was used to determine the actual percentage of the dopant in the lattice as opposed to the nominal doping level.  This was performed using wavelength dispersive x-ray spectroscopy (WDS) in
%the electron probe microanalyzer of
a JEOL JXA-8200 electron-microprobe.  Magnetization data were collected in a Quantum Design (QD) Magnetic Properties Measurement System (MPMS).  Temperature-dependent AC electrical resistance data (f=16Hz, I=3mA) was collected using either a QD MPMS with a LR700 resistance bridge or a QD Physical Properties Measurement System (PPMS). Electrical contact was made to the sample using Epotek H20E silver epoxy to attach Pt wires in a four-probe configuration.

\section{results}

\begin{figure}[h]
	\begin{center}
		\scalebox{0.3}{
		\includegraphics{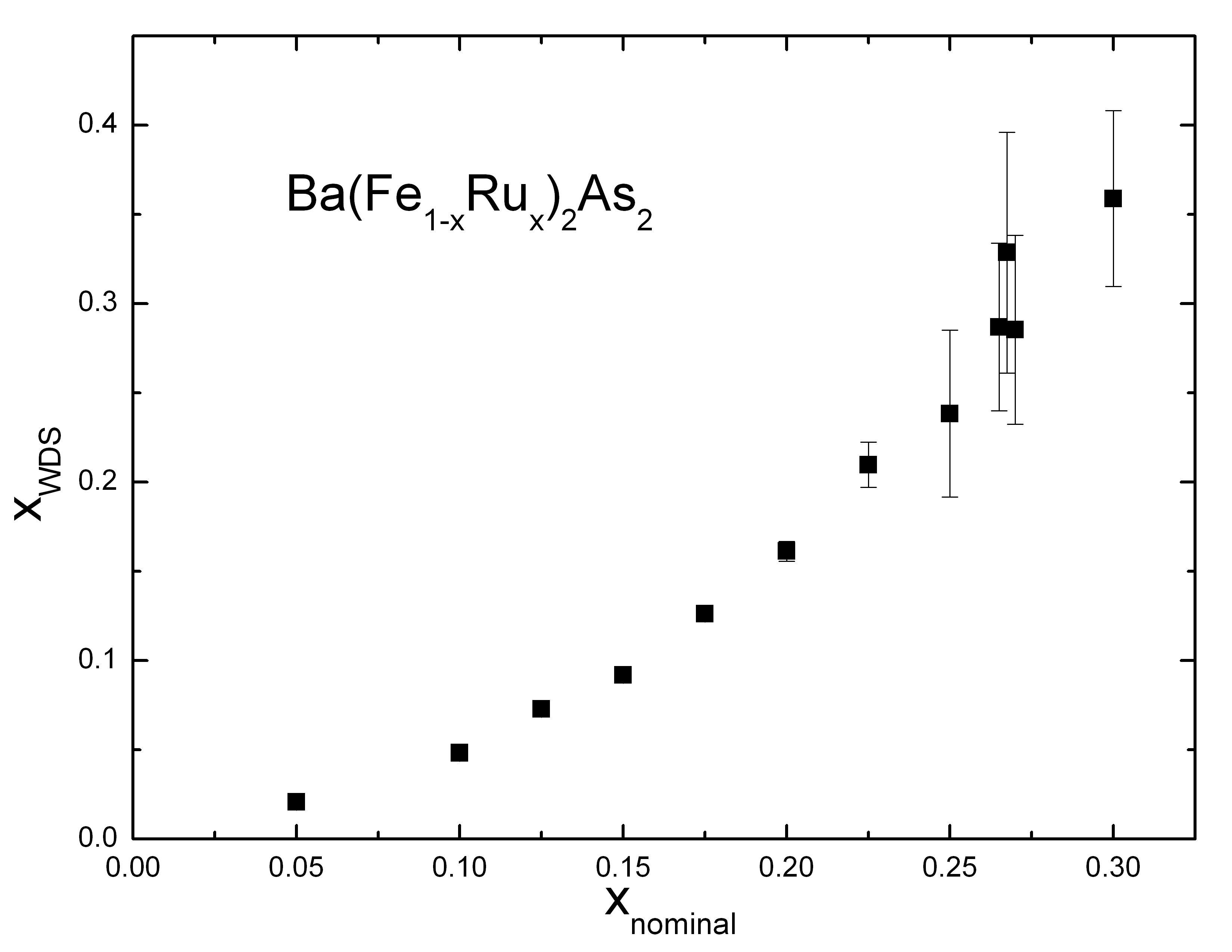}}
	\end{center}
	\caption{Experimentally determined Ru concentration, $x_{WDS}$, vs nominal Ru concentration.  Error bars are $\pm2\sigma$ (values from Table~\ref{table:WDSdata}).}
	\label{fig:rvn}
\end{figure}

A summary of the WDS measurement data is presented in Table~\ref{table:WDSdata}.  For each batch, between 1 and 5 crystal surfaces were measured.  The table shows the number of points measured, the nominal $x$ value measured, the average $x$ value, and two times the standard deviation of the $x$ values measured.  All $x$ values given in this paper are the average $x_{WDS}$ values determined by wavelength dispersive x-ray spectroscopy (WDS).  Fig.~\ref{fig:rvn} shows the measured vs nominal Ru concentration, as well as the error bars on the measured values.  For $x_{WDS}\leq 0.21$ the variation in Ru content within a batch is small, in the range of $1-5\%$ of the $x$ value.  Such variation is similar to what is found for other 3d and 4d doping series\cite{tillman,thaler,canfield3d,canfieldoverview}. For $x\geq 0.24$ there is a sudden and rather dramatic increase in the variation of the Ru concentration within a single batch (and even a single sample).  It is not clear what the origin of the change in homogeneity is, but it is also noted, in a qualitative manner, in ref.\cite{rullier} as well.

\onecolumngrid

\begin{center}
\begin{table}[h]
	\begin{tabular}{c|c|c|c|c|c|c|c|c|c|c|c}
	\hline\hline
	\multicolumn{12}{c}{\baru}\\
	\hline
	N & 14 & 16 & 12 & 12 & 11 & 19 & 18 & 13 & 14 & 15 & 25 \\
	\hline
	$x_{nominal}$ & 0.05 & 0.1 & 0.125 & 0.15 & 0.175 & 0.2 & 0.225 & 0.25 & 0.265 & 0.27 & 0.3 \\
	\hline
	$x_{WDS}$ & 0.021 & 0.048 & 0.073 & 0.092 & 0.126 & 0.161 & 0.210 & 0.24 & 0.29 & 0.29 & 0.36 \\
	\hline
	$2\sigma$ & 0.001 & 0.001 & 0.001 & 0.003 & 0.003 & 0.005 & 0.013 & 0.05 & 0.05 & 0.05 & 0.05 \\
	\hline\hline
	\end{tabular}
	\caption{WDS data for Ba(Fe$_{1-x}$Ru$_x$)$_2$As$_2$.  N is the number of points measured in each batch, $x_{WDS}$ is the average $x$ value for that batch, and $2\sigma$ is twice the standard deviation of the N values measured.}
	\label{table:WDSdata}
\end{table}
\end{center}
%\twocolumngrid

\begin{figure}[h]
%	\scalebox{0.3}{
		\includegraphics[width=.5\linewidth]{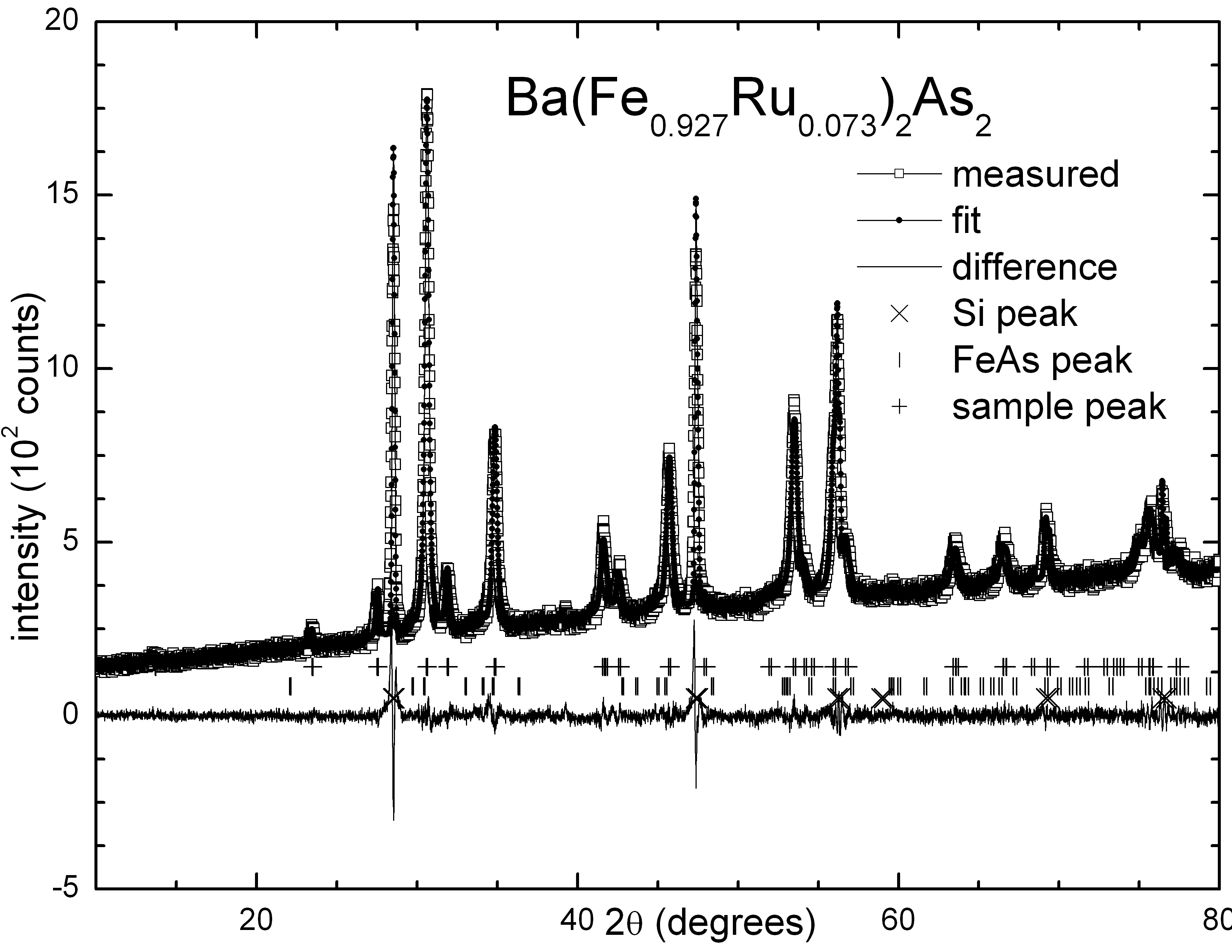}
	\caption{Powder x-ray pattern for Ba(Fe$_{1-x}$Ru$_x$)$_2$As$_2$, $x=0.073$, with Si standard.  Open symbols are measured data, closed ones are fit, the line shows the difference.  $\times$, $|$ and $+$ symbols are calculated peak positions for Si, FeAs and the sample.}
	\label{fig:rawxray}
\end{figure}

\begin{figure}[h]
	\begin{center}
%		\scalebox{0.3}{
		\includegraphics[width=.5\linewidth]{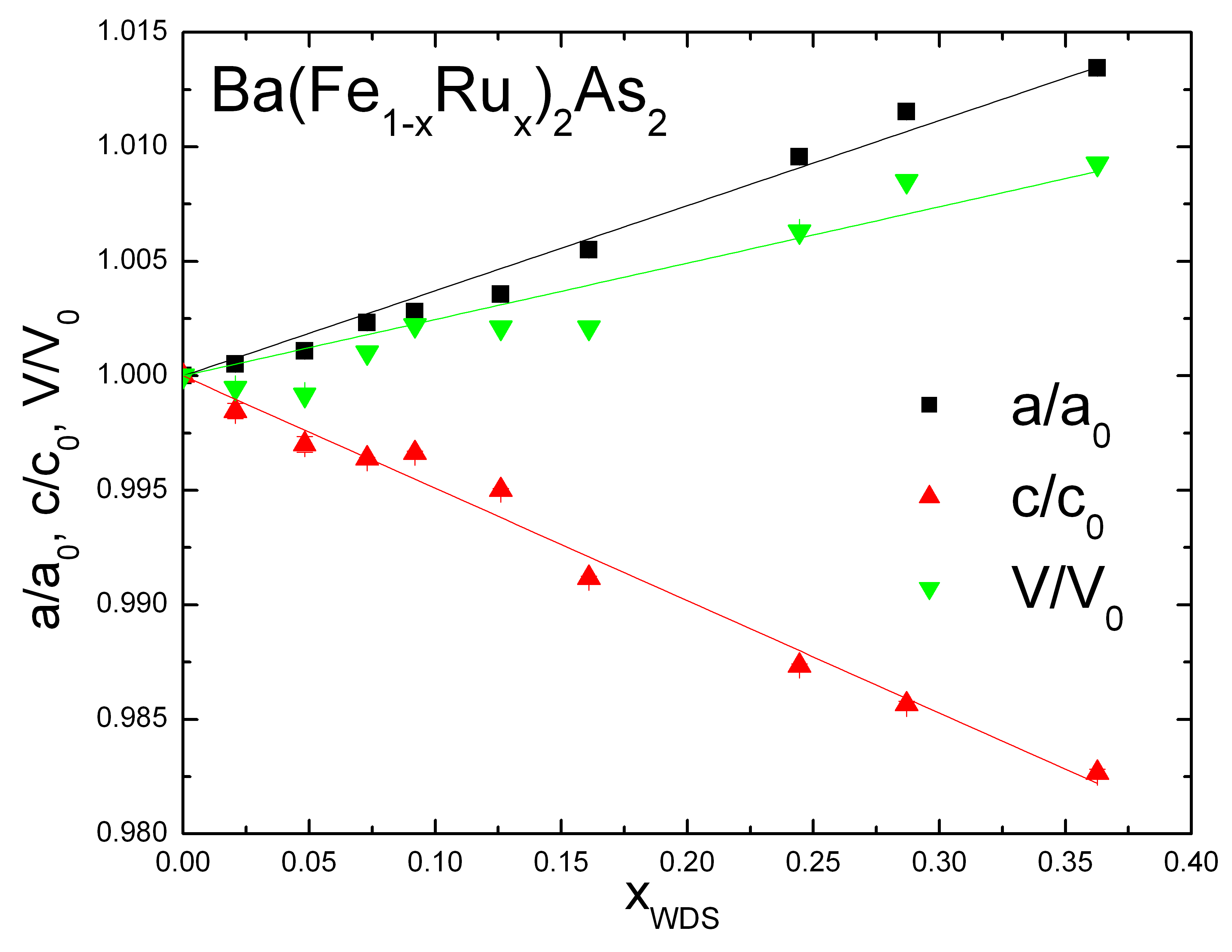}
	\end{center}
	\caption{Lattice parameters for Ba(Fe$_{1-x}$Ru$_x$)$_2$As$_2$, compared to BaFe$_2$As$_2$, for which $a_0=3.96\text{\AA}$, $c_0=13.0\text{\AA}$ and $V_0=204\text{\AA}^3$.  The slopes are $a/a_0: (3.7\pm0.1)\times 10^{-4}{\huge /}\text{Ru atom}$, $c/c_0: (-4.9\pm0.1)\times 10^{-4}{\huge{/}}\text{Ru atom}$, $V/V_0: (2.4\pm0.2)\times 10^{-4}{\huge{/}}\text{Ru atom}$.  The trend lines are determined by a least squares fit.  The error in the slope is the standard error from this fit. (Color online)}
	\label{fig:lpplot}
\end{figure}

Powder x-ray diffraction measurements confirm that \baru forms in the I4/mmm, ThCr$_2$Si$_2$ structure and that impurities are minimal (Fig.~\ref{fig:rawxray}).  Rietveld refinement of the XRD data gives the $a$ and $c$ lattice parameters, which are plotted, along with the unit cell volume, as a function of $x_{WDS}$ in Fig.~\ref{fig:lpplot}.

%The spread in the real Ru concentration becomes large around 22.5\% nominal, 21\% real, and grows from there.

\begin{figure}[h]
	\begin{center}
%		\scalebox{0.3}{
		\includegraphics[width=.5\linewidth]{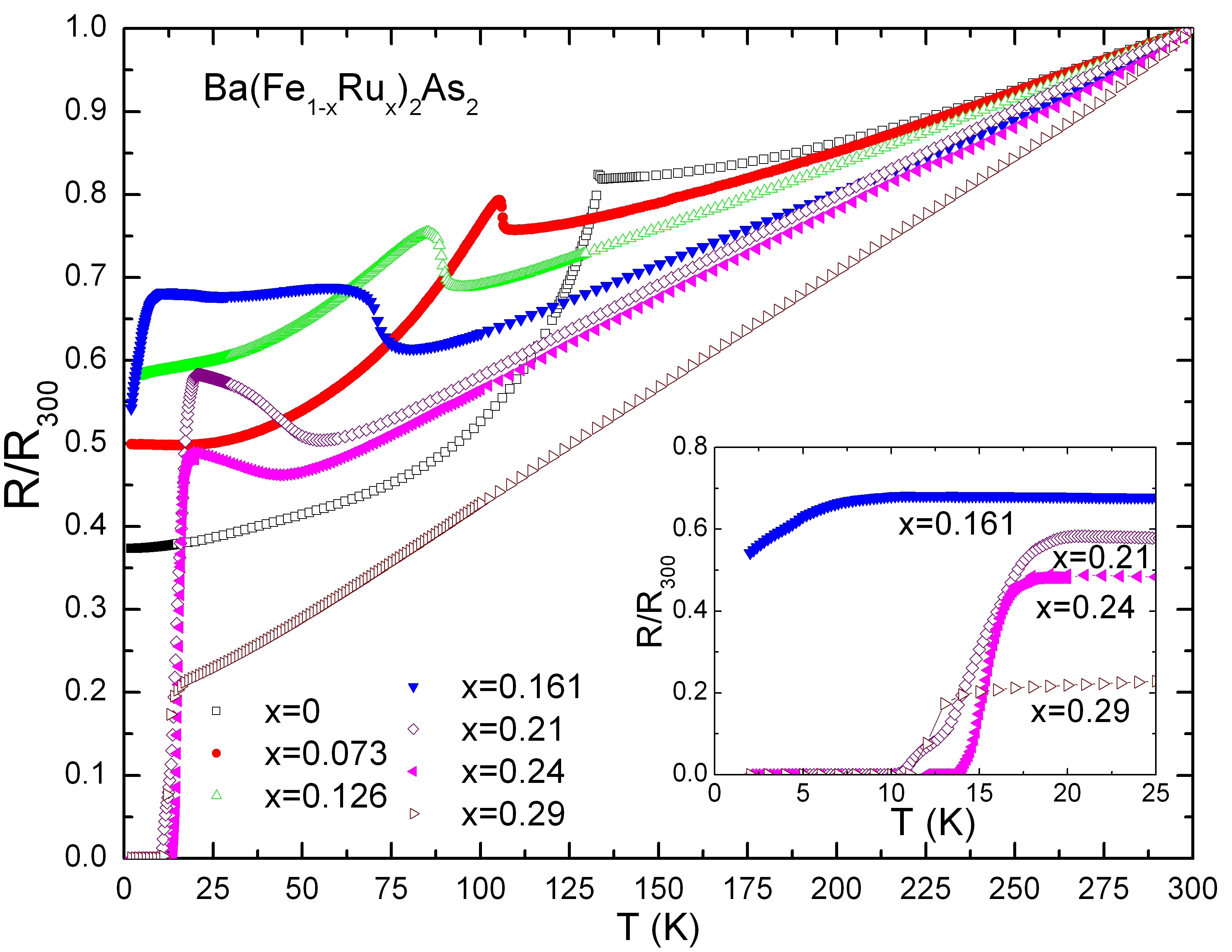}
	\end{center}
	\caption{Temperature dependent resistance, normalized to the room temperature value, for select Ba(Fe$_{1-x}$Ru$_x$)$_2$As$_2$ doping levels.  Inset shows low temperature behavior. (Color online)}
	\label{fig:RRRdata}
\end{figure}

Figure~\ref{fig:RRRdata} shows the normalized electrical resistance data of the \baru series from 5~K to 300~K.  Normalized resistance is plotted instead of resistivity because of the tendency of these samples to exfoliate or crack\cite{tillman,tanatar:094507,tanatar:134528}.  The anomaly in normalized resistance at 134~K for pure BaFe$_2$As$_2$ is
%explained by the structural/magnetic phase transitions.
associated with a first order phase transition into an orthorhombic antiferromagnetic state\cite{rotter:020503}.  As in the case of Co, Ni, Cu, Rh and Pd substitution\cite{tillman,canfield3d,thaler,canfieldoverview}, the temperature of the resistive anomaly is suppressed monotonically and the shape is changed from the sharp loss of resistance on cooling through $T_S/T_m$ seen in pure BaFe$_2$As$_2$ to a broader increase in resistance on cooling through $T_S/T_m$ for intermediate $x$ values.  For $x\geq0.29$, anomalies associated with $T_S/T_m$ are no longer detectable.  Superconductivity begins to appear above $x=0.161$ (resistive onset only) and is fully manifested ($R=0$) by $x=0.210$.  A maximum $T_c$ of 16.5~K is achieved at $x\approx0.29$.  $T_c$ is suppressed for higher values of $x$.  The superconducting transition is quite broad compared to other TM dopings: more than 7~K wide for $x_{WDS}=0.210$ compared with a 3~K width for a Co doping level of $x=0.038$\cite{tillman}.  Such a wide transition is more typical of pressure induced superconductivity rather than chemical doping\cite{colombier}.

\begin{figure}[h]
	\begin{center}
%		\scalebox{0.3}{
		\includegraphics[width=.5\linewidth]{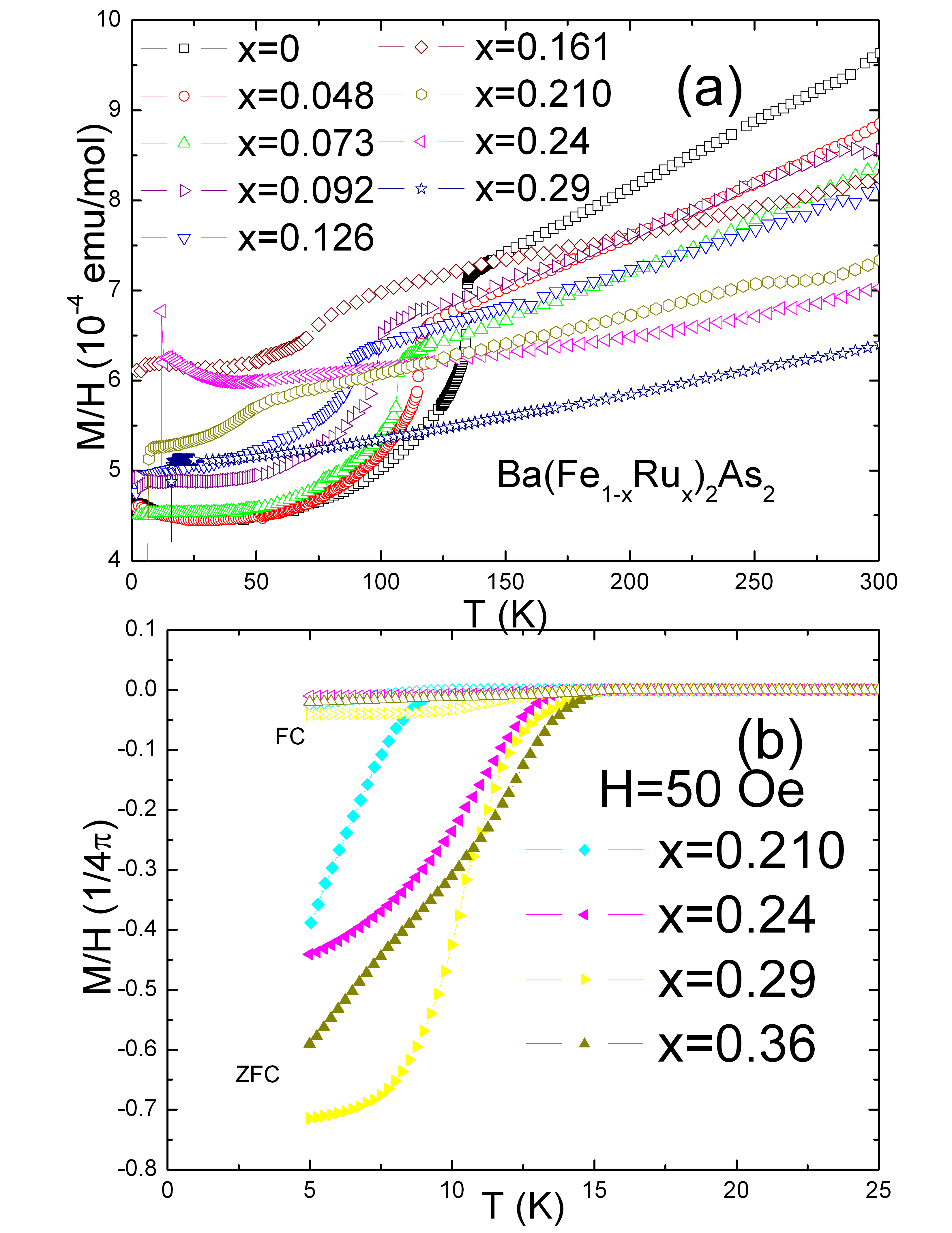}
	\end{center}
	\caption{(a) Temperature dependent magnetization, scaled by applied field $H=70$ kOe, for Ba(Fe$_{1-x}$Ru$_x$)$_2$As$_2$. (b) Low temperature, low field, zero field cooled and field cooled magnetization for several superconducting members of the \baru family.  $H\perp c$ for all data sets.  The relatively large, low temperature, diamagnetic shielding in the zero field cooled measurements approaches that found for Co, Ni, Rh and Pd doping\cite{thaler,tillman,canfield3d,canfieldoverview}.}
	\label{fig:magdata}
\end{figure}

Figure~\ref{fig:magdata}(a) shows high field (H=70 kOe) $M/H$ data for representative members of the \baru series.
%$H$ is perpendicular to $c$ in these data sets.
%It is clear here that the high temperature magnetic signature is weaker in the higher dopings.
At high temperatures the $M(T)/H$ ratio is roughly linear and decreases with decreasing temperature, with a slope that decreases with increasing Ru doping.
%The structura/magnetic phase transition can be seen to be suppressed until it totally vanishes in the highest doping levels.
As with normalized resistance, the magnetization of the parent compound manifests a clear change at 134~K, correlated with the structural/magnetic phase transition\cite{rotter:020503}.  As $x$ is increased up to $x=0.126$, this transition is suppressed and broadened without qualitative change.  Starting with $x=0.161$ the transition becomes much flatter and broader, and by $x=0.24$ it is barely visible.  At $x=0.29$ it has completely vanished.

Figure~\ref{fig:magdata}(b) shows the low field (50 Oe) $M/H$ data for the superconducting members of the \baru series.
%These data were taken with $H$ perpendicular to the $c$ axis and with H=50 Oe.
These samples show a clear diamagnetic signal in the zero field cooled (ZFC) data,
%  The field cooled (FC) data shows a much lower diamagnetic signal, suggesting that Meissner expulsion is weak in this material.\cite{thaler}
as well as some Meissner expulsion.  It is worth noting that whereas the ZFC diamagnetic signal for Co, Ni, Rh, Pd and Cu/Co dopings are all similar and close to $-1/4\pi$\cite{tillman,thaler,canfield3d,canfieldoverview,nithesis}, the low temperature values for Ru doping (\ref{fig:magdata}(b)) are smaller in amplitude and vary more.

%Heat capacity

\section{discussion}
\begin{figure}[h]
	\begin{center}
%		\scalebox{0.35}{
		\includegraphics[width=.55\linewidth]{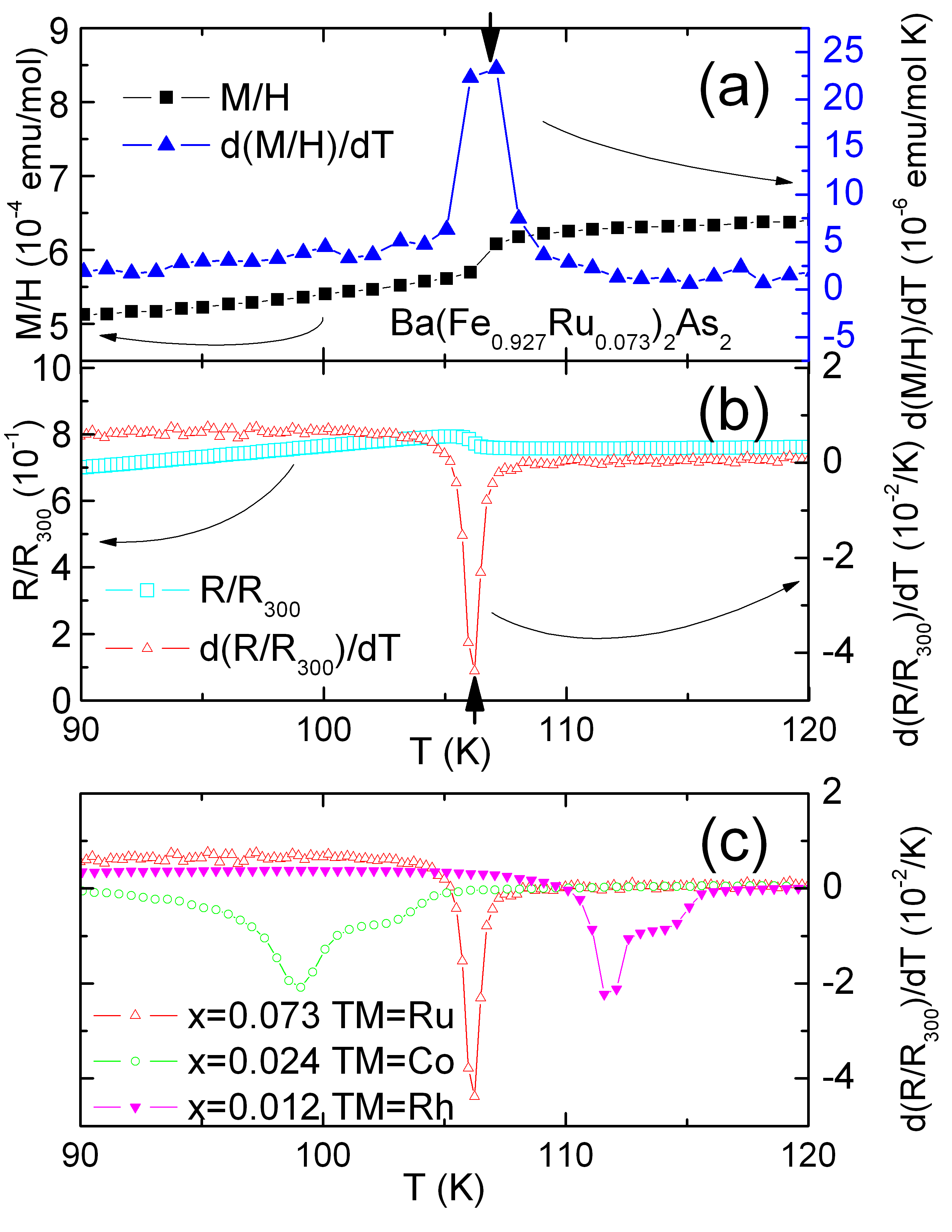}
	\end{center}
	\caption{Magnetization (a) and normalized resistance (b), along with derivatives, for \baru ($x=0.073$).  Vertical arrows show the criteria for determination of the transition temperature.  (c) shows normalized resistance derivative data for Co doping (x=0.024) and Rh doping (x=0.012) with similar transition temperatures. (Color online)}
	\label{fig:kq918transplot}
\end{figure}

\begin{figure}[h]
	\begin{center}
%		\scalebox{0.3}{
		\includegraphics[width=.5\linewidth]{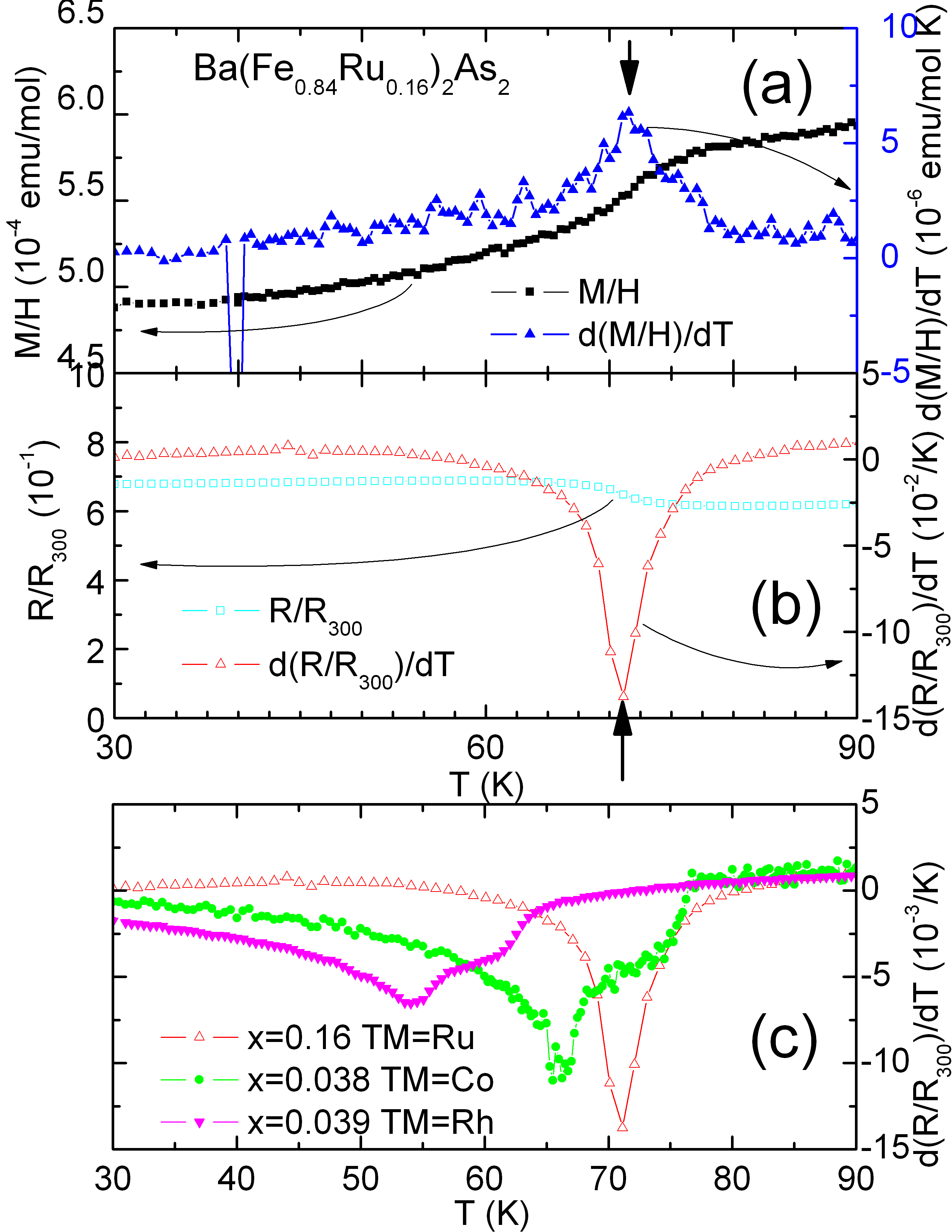}
	\end{center}
	\caption{Magnetization (a) and normalized resistance (b), along with derivatives, for \baru ($x=0.16$).  Vertical arrows show the criteria for determination of the transition temperature.  (c) shows normalized resistance derivative data for Co doping (x=0.038) and Rh doping (x=0.039) with similar transition temperatures. (Color online)}
	\label{fig:kq355transplot}
\end{figure}

Figures~\ref{fig:kq918transplot}~and~\ref{fig:kq355transplot} show normalized resistance and magnetization data, along with their derivatives, for $x=0.073$ and $x=0.16$ samples respectively.  These figures show the criteria used for determining the structural/magnetic phase transition temperatures for these materials.
%The positive spike in the derivative of magnetization and negative spike in the derivative of normalized resistance correspond to the phase transition temperature.

%The wider spike in the more highly doped sample shows that the transition is becoming broader as x increases.  In other superconduting members of the Ba(Fe$_{1-x}$TM$_{x}$)$_2$As$_2$ family, the structural and magnetic transitions have been shown to split at higher dopings\cite{thaler,tillman,pratt}.  In addition to magnetic vs resistive temperature differences, this splitting has a definite signature in the resistive and magnetic data.  In this system, we do not see an appreciable difference in the temperature difference of the magnetic and resistive transition signatures (see Fig.~\ref{fig:kq918transplot}(c) and \ref{fig:kq355transplot}(c)).

Figures~\ref{fig:kq918transplot}(c)~and~\ref{fig:kq355transplot}(c) show comparisons of normalized resistance derivatives for Ru, Co and Rh doped BaFe$_2$As$_2$ with similar $T_S/T_m$ values.
%In all three systems, the wider spike shows that the transition broadens at higher doping.
In the Co and Rh series,
%display both a minimum and a "shoulder" in the derivative.
a clear splitting of the two transitions is visible.  (At the same temperatures, the derivatives of magnetization and heat capacity show split features as well\cite{tillman,thaler,canfield3d,canfieldoverview}.)  By contrast, we do not see these separated features in the derivatives of the normalized resistance from the Ru system.  These features have been shown to correspond to a splitting of the joint transition into two transitions, one structural the other magnetic\cite{thaler,canfield3d,pratt,canfieldoverview,RhDistort}.  Although the authors of ref.\cite{rullier} claim to see a split transition, it appears to be a subtle feature compared to Co or Rh data.  The single feature in the Ru doped series $dR/dT$ data suggests that either the splitting is much smaller, or absent, in this system or that the resistive feature associated with $T_S$ is much weaker in this system.  It is possible that the splitting is caused by the extra electrons provided by other TM doping (eg. Co, Ni, Cu, Rh, Pd).

Onset and offset criteria were used to determine $T_c$ from this resistance data.  $T_c$ was determined from the magnetization data by extrapolating the maximum slope of the ZFC data back to the normal state.  There is fair agreement between $T_c^{offset}$ determined from normalized resistance and $T_c$ determined from magnetization.  It should be noted, though, that (i)~superconductivity primarily occurs in the region where the spread in $x_{WDS}$ is large, and (ii)~the superconducting transition is broad in $R(T)$ and both ZFC and field cooled Meissner data are somewhat lower than for other TM doped series.
%what do I say about this?

\begin{figure}[h]
	\begin{center}
%		\scalebox{0.3}{
		\includegraphics[width=.5\linewidth]{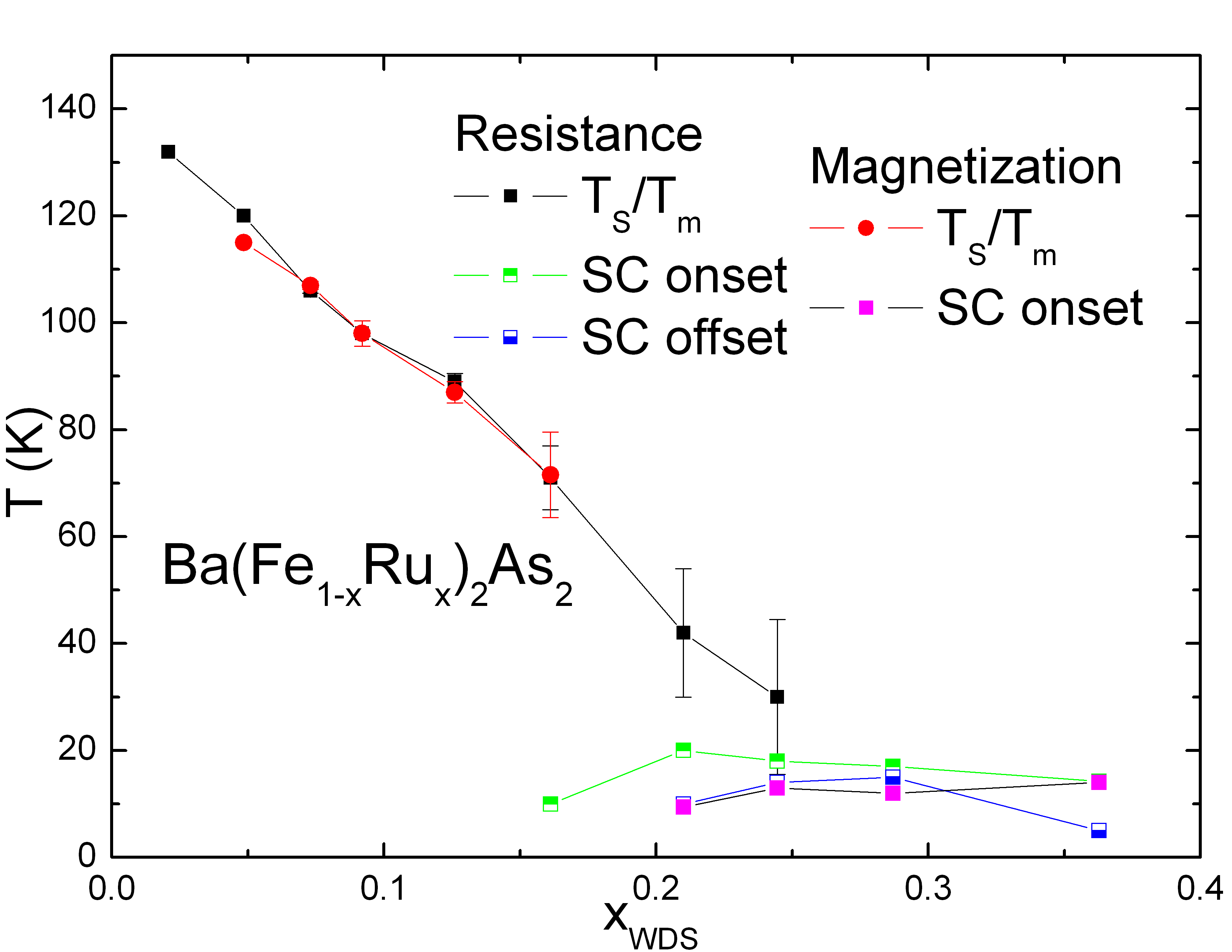}
	\end{center}
	\caption{$x$ dependent phase diagram, showing $T$ for salient features in Ba(Fe$_{1-x}$Ru$_x$)$_2$As$_2$.  (Color online)}
	\label{fig:pdcomplete}
\end{figure}

\begin{figure}[h]
	\begin{center}
%		\scalebox{0.3}{
		\includegraphics[width=.5\linewidth]{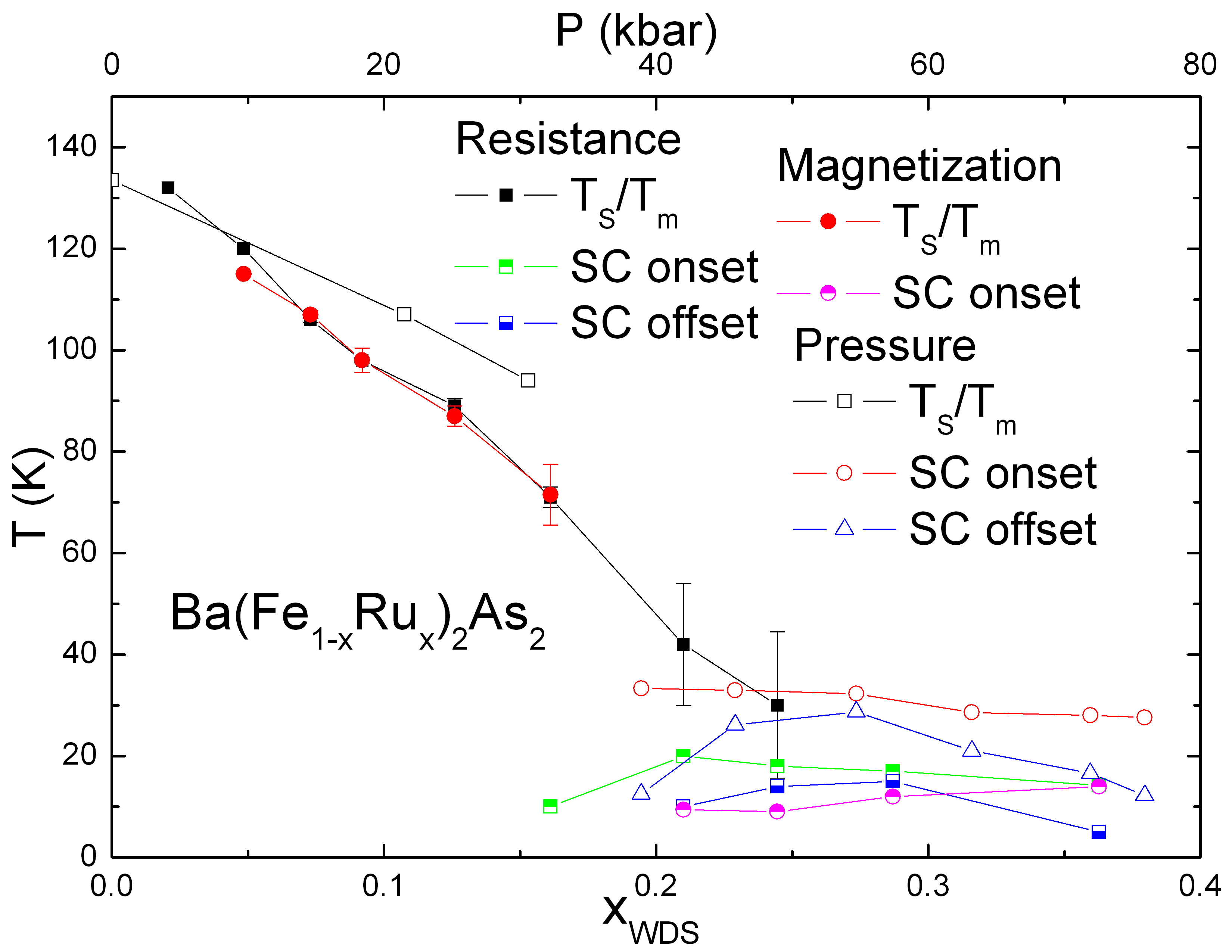}
	\end{center}
	\caption{Comparison of Ru doping phase diagram with that of the parent BaFe$_2$As$_2$ compound under applied pressure.  (Color online)}
	\label{fig:pdcompletewp}
\end{figure}

Using these criteria, the data presented in Figs.~\ref{fig:RRRdata}~and~\ref{fig:magdata} are summarized in a $T-x$ phase diagram shown in Fig.~\ref{fig:pdcomplete}.
%As can be seen in the phase diagram, the higher temperature structural and magnetic phase transitions are suppressed in a similar manner as with other dopants\cite{tillman,chu,ning,fang,lester,pratt,canfield3d,thaler,wang,saha}.  As with other dopants, superconductivity is found in both tetragonal and orthorhombic phases and is stable over a dome-like region\cite{tillman,chu,ning,fang}.  The maximum $T_c$ is lower here than for other $TM$ dopings, with a maximum of about $16.5K$.
Overall, the phase diagram for the \baru series is qualitatively quite similar to that of the Co, Ni, Rh and Pd diagrams: increasing $x$ suppresses the structural/magnetic phase transition, a superconducting dome appears above some critical $x$ value, and this dome has a maximum near the point where $T_S/T_m$ extrapolates to zero.  However, there is a key difference: suppression of $T_S/T_m$ is much slower than for other TM dopings (Co, Ni, Cu, Rh, Pd)\cite{tillman,thaler,canfield3d,canfieldoverview,nithesis}.  In previous comparisons of 3d and 4d TM dopings\cite{thaler,canfield3d,canfieldoverview}, we showed that suppression of $T_S/T_m$ occurs at roughly the same rate regardless of differences in size and electron count between dopants;
%Also, the wide superconducting transitions in \baru are not seen in the other TM dopings.
the suppression of $T_S/T_m$ in \baru is about three times slower.

As in the case of Rh and Pd doped BaFe$_2$As$_2$\cite{thaler}, with Ru doping the $c$-lattice parameter shrinks compared to the parent BaFe$_2$As$_2$, while the $a$-lattice parameter and the unit cell volume, V, grow.  (This is in contrast to the 3d TM dopings, where all three shrink with increasing $x$.)
%At $x=0.245$, $a=1.007a_0$ and $c=0.99c_0$, which is comparable to the trend seen in polycrystalline doping\cite{sharma} as well as other recent single crystal work.\cite{rullier}
By way of comparison: a Ru doping level of $x_{WDS}=0.175$ has $a=1.002a_0$ and $c=0.995c_0$ and a Rh doping level of $x=0.171$ has $a=1.007a_0$ and $c=0.988c_0$.\cite{thaler}
%This is different from the trend seen between Rh and Pd.  When comparing Rh with equivalent Pd dopings, $a$ is slightly larger and $c$ is slightly smaller in the Rh doping\cite{thaler}.  Ru doping giving smaller values for $a$ and larger ones for $c$ is an oddity which may provide insight into the steric contributions to the behavior of this system.
Because the crystallographic trends of all three 4d TM dopant series (Ru, Rh and Pd) are similar, the major differences in their $T-x$ phase diagrams suggest that steric effects alone are not enough to explain the differences in behavior of this system with doping (ie. the extra electrons in Rh and Pd are responsible for the much more rapid effects of doping).

\begin{figure}[h]
	\begin{center}
%		\scalebox{0.3}{
		\includegraphics[width=.5\linewidth]{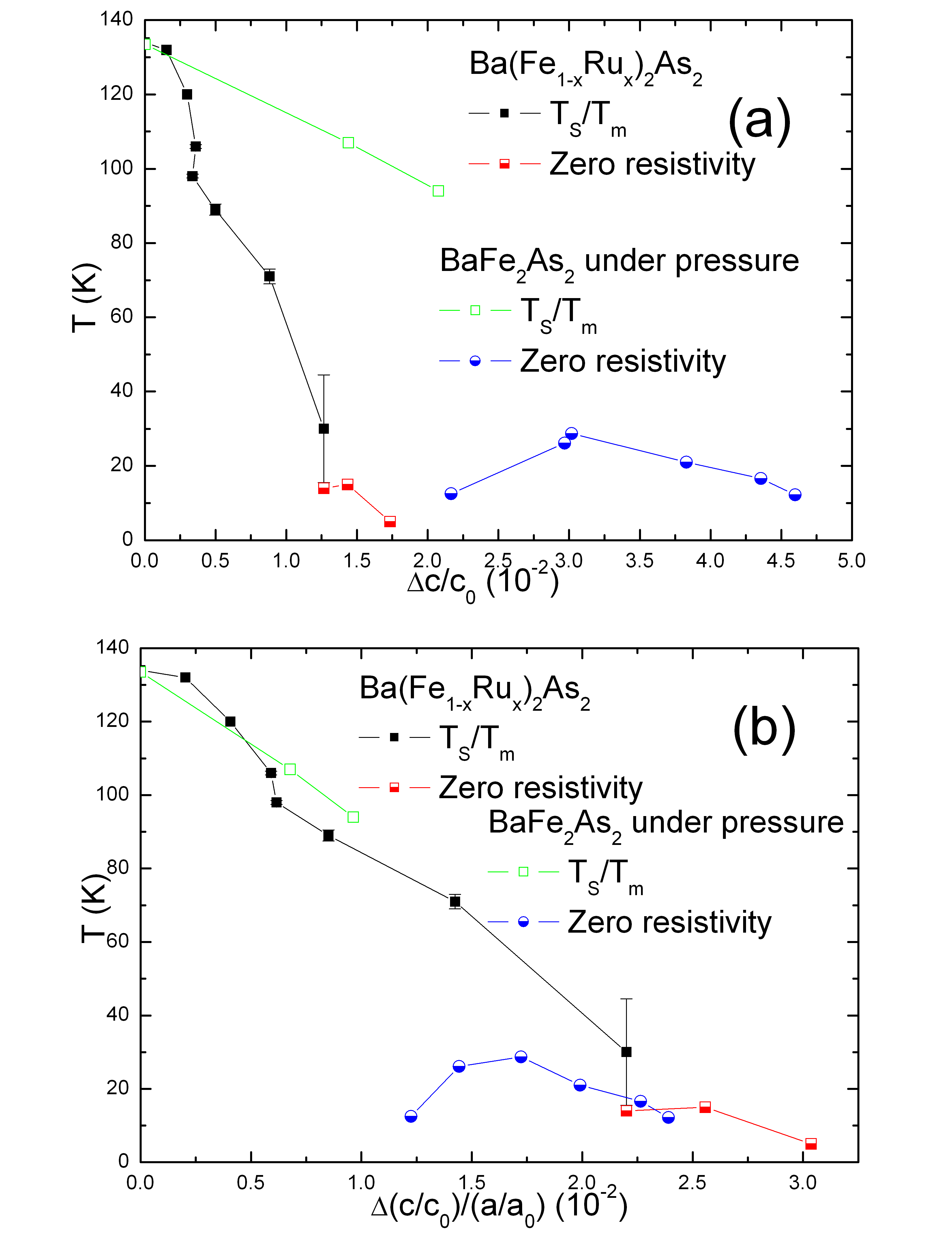}
	\end{center}
	\caption{Phase diagrams of \baru and of parent BaFe$_2$As$_2$ under pressure, scaled by lattice parameters.  (a) is scaled by $\Delta c/c_0$.  (b) is scaled by  $\Delta(c/c_0)/(a/a_0)$. (Color online)}
	\label{fig:LPscaledPDplot}
\end{figure}

Although the maximum superconducting critical temperature, $T_c^{max}$, is significantly lower in the Ru doped system, there is a clear similarity between the Ru doped $T-x$ phase diagram and the pressure dependent, $T-p$, phase diagram of the parent BaFe$_2$As$_2$ compound\cite{colombier}, as can be seen in Fig.~\ref{fig:pdcompletewp}.
%In both cases, the superconducting transition is wide compared with the electron side TM dopings.
The similarity of the phase diagrams suggests that changes in the unit cell dimensions may be playing a large role in determining the superconducting behavior, with the effects of Ru substituion in this system being similar to physical pressure in the undoped BaFe$_2$As$_2$ system.  The difference in $T_c^{max}$ is most likely caused by the Ru dopant disordering the Fe-plane, whereas pressure induces no such distortion.
%Since the $c$ lattice parameter shrinks in both cases, we can use it to guide scaling of the phase diagram when comparing Ru doping with pressure application.  Fig.~\ref{fig:LPscaledPDplot} shows two different scaling comparisons of the \baru phase diagram with that of pressure. Fig.~\ref{fig:LPscaledPDplot}(a) was constructed by matching the raw c lattice parameter and scaling the phase diagram to match.  Fig.~\ref{fig:LPscaledPDplot}(b) matches the ratio of $(c/c_0)/(a/a_0)$ and uses this to scale instead.  As shown in Fig.~\ref{fig:LPscaledPDplot}(a), raw scaling by $c$ does not work, but as Fig.~\ref{fig:LPscaledPDplot}(b) demonstrates, we see that the progression of $T_S/T_m$ suppression is very close and the superconducting domes overlap when scaling by $\Delta(c/c_0)/(a/a_0)$\cite{kimber}.

Whereas the agreement between the $T-x$ and $T-p$ phase diagrams in Fig.~\ref{fig:pdcompletewp} is good, the scaling between $x$ and $p$ was arbitrarily choosen to optimize the overlap of the two data sets.  Using our data on the $x$-dependence of the unit cell parameters (\ref{fig:lpplot}) in combination with the data from ref.~\cite{kimber} on the pressure dependence of the unit cell parameters of BaFe$_2$As$_2$, we can make this comparison more quantitative.  Of the four combinations of the unit cell parameters: $a$, $c$, V and $c/a$, only $c$ and $c/a$ show similar responses to pressure and doping; $a$ and V both increase with doping whereas they decrease with $p$.  Figures~\ref{fig:LPscaledPDplot}(a) and (b) present our Ru-doping data as well as the pressure data from ref.~\cite{colombier} plotted as functions of the changes in $c$ and $c/a$.  A comparison of these two figures clearly indicates that $c/a$ rather than $c$ better parameterizes the effects of doping and pressure.  This result means that, based on {\it these two} isoelectronic perturbations (pressure and Ru doping), changes in the $c/a$ ratio appear to be more physically important than changes in $c$ alone.

The other isoelectronic substitution which produces superconductivity in BaFe$_2$As$_2$ is P doping on the As site\cite{jiang,kasahara}.  Although the maximum $T_c$ in the BaFe$_2$(As$_{1-x}$P$_x$)$_2$ system is quite a bit higher than in the \baru system ($\sim30$K), several key properties are similar.  $T_S/T_m$ is suppressed in a relatively gradual manner and the maximum $T_c$ value occurs at a comparably high doping level ($x_{Ru}=0.29$, $x_P=0.32$) and extends over a much wider range than in any of the electron doped TM series\cite{jiang,kasahara}.  Furthermore, both Ru doping and underdoping of P produce wider transitions than other TM dopings (eg. Co, Ni, Rh, Pd)\cite{tillman,li,thaler,kasahara}.
%Finally, the $c$ lattice parameter shrinks with increasing $x$ and $c/c_0$ for the maximum T$_c$ are similar: 0.985 for P and 0.986 for Ru\cite{jiang}.  This is in contrast to Co and Rh, where $c/c_0$ for maximum T$_c$ are 0.997 and 0.996 respectively\cite{thaler}.
On the other hand, taking changes in $c$ and $a$ with P doping into account, $T_S/T_m$ and $T_c$ for P-doped and Ru-doped BaFe$_2$As$_2$ scale better with changes in $c$ than with changes in $c/a$\cite{kasahara}.  This means that, if we include P-doping as a third isoelectronic perturbation, then neither changes in $c$ nor $c/a$ universally describe the $T-x$ and $T-p$ phase diagrams.

%During the final stages of this work, we have learned of transport and electronic structure measurements recently done on the \baru system\cite{rullier,brouet}.  Data presented by Rullier-Albenque et al. is qualitatively quite similar to what is shown here\cite{rullier}.  However, the specifics are somewhat different.  In particular, max T$_c$ is reported to occur at $x=0.35$ versus $x=0.29$ for this work, and the absolute maximum differs by about 3K.  Further work is obviously needed.  Brouet et al. have demonstrated that isovalent Ru substitution produces effects in the Fermi surface that are similar to K or Co doping\cite{brouet}.  This suggests that the mechanism by which superconductivity is induced in this system is the same as for other dopings.

\section{Summary}
Single crystals of \baru can be grown for $x<0.37$, although Ru homogeneity becomes less well controlled for $x>0.21$.  The structural and magnetic phase transition temperature, $T_S/T_m$, is suppressed as $x$ increases but does not clearly split, as it does for TM = Co, Ni, Cu, Rh, and Pd doping.  As $T_S/T_m$ is suppressed superconductivity appears, reaching a maximum $T_c$ value of 16.5~K for $x = 0.29$, near the point that $T_S/T_m$ extrapolates to $T = 0$~K.  Whereas the suppression of $T_S/T_m$ and the stabilization of $T_c$ occur at a much slower rate for Ru doping than they do for doping with TM = Co, Ni, Cu, Rh, or Pd, indicating that the additional electrons brought by these dopants play a significant role in tuning of this system, there is a remarkable agreement between two isoelectronic phase diagrams (Ru-doping and pressure) of BaFe$_2$As$_2$ when plotted as $T(c/a)$, but not when plotted as $T(c)$.

%\section{Conclusions}
%Ru substitution on the Fe site in BaFe$_2$As$_2$ suppresses the structural/magnetic phase transition and induces superconductivity.  Homogeneous doping is difficult to achieve for higher doping levels, and superconductivity appears primarily in this inhomogeneous region.  The superconducting transition is broad and "reduced" in $\chi(T)$.

%Several questions are left open for the future. Whether this system is superconducting in the bulk in the same way as in electron side TM doping is still worth investigation.  Are the carriers provided by electron side TM doping important for splitting the structural/magnetic transition, and is this related in some way to the much higher doping level required by isovalent doping?

\section{Acknowledgements}
Work at the Ames Laboratory was supported by the Department of Energy, Basic Energy Sciences under Contract No. DE-AC02-07CH11358. We would like to thank E. D. Mun, S. Kim, X. Lin, R. Gordon, R. Fernandes, A. Kreyssig, R. Roggers and J. Brgoch for help and useful discussions.

\bibliography{rupaper}
\end{document}